\documentclass[11pt]{article}
\usepackage{moriond,epsfig}

\def\Journal#1#2#3#4{{#1} {\bf #2}, #3 (#4)}

\def\AandA{\em A\&A}
\def\APJ{\em ApJ}
\def\MN{\em MNRAS}

\def\PASP{\em PASP}

\def\be{\begin{equation}}
\def\ee{\end{equation}}
\def\bea{\begin{eqnarray}}
\def\eea{\end{eqnarray}}

\def\lya{Ly$\alpha$}
\def\ha{H$\alpha$}
\def\bet{$\beta$}
\def\bmap{$\beta$-map}
\def\eso338{ESO\,338-IG04}
\def\fluxunit{erg~s$^{-1}$~cm$^{-2}$}

\def\fluxiuee338{$123\times10^{-14}$\fluxunit}
\def\fluxacse338{$134\times10^{-14}$\fluxunit}
\def\wlya{$W({\rm Ly}\alpha)$}

\begin{document}
\vspace*{4cm}
\title{Lyman $\alpha$ emission from local starburst galaxies and high-$z$ 
objects}

\author{ G. \"Ostlin (1), M. Hayes (1), J.M. Mas-Hesse (2), 
D. Kunth (3), C. Leitherer (4) \& A. Petrosian (5) }

\address{ (1)Stockholm Observatory, AlbaNova University Centre, 
          106 91 Stockholm, Sweden\\
          (2) Centro de Astrobiolog\'{\i}a (CSIC--INTA), 
          Madrid, Spain\\
          (3) Institut d'Astrophysique, Paris (IAP), 
          France\\
          (4) Space Telescope Science Institute, 
          Baltimore, MD 21218, USA\\
          (5) Byurakan Astrophysical Observatory and 
          Isaac Newton Institute of Chile, 
          Armenia }

\maketitle\abstracts{
We review the history and the current status of the 
understanding of the processes that regulate Lyman$\alpha$ emission from 
star-forming galaxies.
 We present some of the most recent results of our study to image 
{\em local} starburst galaxies in the \lya\ emission line using the Advanced
Camera for Surveys on the Hubble Space Telescope. 
 Particular attention is dedicated to our study of the low-metallicity, 
dust-poor Blue Compact Galaxy \eso338. 
 We discuss some of our local observational results with reference to the 
interpretation of results of high-redshift \lya\ surveys and recent 
simulations of the detection properties of high-$z$ \lya\ emitting objects
performed by our group.	 }

\section{Motivation for Lyman$\alpha$ studies -- the current status\label{sect:intro}}

 Naively, one could expect \lya\ to be the strongest emission line observed 
from hydrogen recombination nebulae with 
$j_{{\rm Ly}\alpha}/j_{{\rm H}\alpha} = 11$ for case A recombination. 
 The potential importance of this line in tracing galaxy formation and 
evolution in the high-redshift ($z$) was postulated in 1967 \cite{pp} 
when it was predicted that the \lya\ luminosity could amount to a few 
percent of the total output.
 It was later predicted that \lya\ equivalent widths could exceed 100\AA\ 
for young, dust-free starbursts \cite{cf}. 
      
 There are numerous methods by which we can study distant galaxies in the 
context of astrophysical cosmology. 
 Tracers such as restframe ultraviolet (UV) continuum, supernovae and assorted
emission lines are often used to study star-formation rates (SFR), angular 
correlation functions, galaxy clustering, and reionisation. 
 Each of these tracers may have their own advantages and disadvantages. 
 Rest-frame UV or colour surveys will be biased towards galaxies with the 
brightest continua. 
 In the hierarchical structure formation model, the more massive, brighter
galaxies are built from more numerous but less massive and fainter ones. 
 The small galaxies, that occupy the faint end of the luminosity function 
will be missed by surveys that target restframe UV continuum. 
 In contrast surveys that target emission lines will not exhibit such a bias.
 A responsive and frequently used tracer of star-formation in 
galaxies is the \ha\ emission line whose equivalent width in the 
absence of dust is 
proportional to the SFR \cite{ken} \cite{mad}.
 However the \ha\ line is redshifted out of the $K$-band at $z\sim2.5$ to  
a spectral domain crowded by sky-lines, rendering \ha\ a sub-optimal
tracer of the very distant universe in ground-based surveys. 
 In contrast, one of the main advantages of the \lya\ line is that it is 
observable in the optical domain to $z\sim5.5$ and in the near infrared to 
$z \sim 17$ I.e. into the suspected neutral universe, before reionisation.

 The first early searches for high-$z$ primeval \lya\ emitters were largely
fruitless, finding number densities far lower than predicted \cite{prit}.
 This  was apparently confirmed by  local \lya\ 
studies \cite{gia}\cite{cf}: \lya\ emitters were scarce and less 
luminous than expected when detections were made. 
 The explanation for the absence of \lya\ galaxies was first thought to be 
due to resonant scatterings H{\sc i} local to the recombination 
nebulae \cite{neu}.
 Resonant scattering leads to an increased probability of \lya\ destruction
either by interactions with dust grains or the two-photon spontaneous channel.
 If the absorbing material is static, column densities of
$\sim 10^{19}$cm$^{-2}$ should be enough to produce damped absorption.
 While the \lya--metallicity relationship \cite{cf} may maintain its 
validity, there are a number of other complicating factors that surround
\lya\ and recent observations using the Hubble Space Telescope (HST) have 
shed light on the real complexity of the issue. 
 Eight nearby star-forming galaxies have been studied by members of our group
using the Goddard High Resolution Spectrograph (GHRS) on HST 
\cite{kun1}: 4 of the sample showed damped absorption, regardless of 
their dust content (including {\sc i}Zw\,18, the most metal-poor 
galaxy known); while the remaining 4 exhibited clear \lya\ 
P-Cygni profiles, characteristic of a large-scale outflow of the 
absorbing gas.
 Subsequent HST observations using the Space Telescope Imaging Spectrograph 
(STIS) \cite{mas} have identified large shells of expanding neutral gas 
covering areas much larger than the starburst. 
 The situation is further complicated by the suggestion that \lya\ emission
from starburst knots could also be geometry-dependent \cite{tt}.
 In this scenario, we could expect to see \lya\ emission only if we 
look directly into an ionised cone created by the starburst; observations
from other angles should observe neutral hydrogen and \lya\ absorption.
 From spectroscopic studies of local starbursts, we can identify several 
mechanisms by which \lya\ photons can escape trapping by 
H{\sc i} columns: emission through P-Cygni profiles in which the H{\sc i}
is outflowing sufficiently to reduce the {\em neutral} column density; 
and escape through a porous medium in which the H{\sc i} covering factor 
is less than unity.
 Recently, our imaging project (Sect. \ref{sect:imaging}) has also 
found evidence for escape {\em through} H{\sc i} regions with column 
densities great enough to produce damped absorption.
 Here photons escape by either scattering into the wing of the emission 
profile or diffusing their way into optically thin regions.

 More recently, detections of high-$z$ \lya\ galaxies have become common
with narrowband searches yielding many sources 
\cite{ouchi} \cite{mr} \cite{fynbo}.
 The high-$z$ situation with regard to \lya\ equivalent width is, however, 
somewhat reversed with surveys finding median $W_{\rm rest}(\rm Ly\alpha)$
of up to 400\AA. 
 However if the high-$z$ results are to be trusted in the cosmological
context, \lya\ needs to be 
calibrated and a rigorous understanding of the statistics is required. 
 Moreover, we need to fully understand any observational effects or 
astrophysical phenomena that can bias such surveys.

\section{\lya\ emission from local starbursts -- the imaging approach\label{sect:imaging}}

 Because of the radiative transfer effects involved, the various possible
escape mechanisms for \lya\ photons are highly dependent upon the ISM 
configuration and an imaging survey with high spatial resolution is essential. 
 This was the motivation for our study to image six local starbursts
with the Solar Blind Channel (SBC) of the Advanced Camera for Surveys on HST.
 In this study (GO program 9470) the {\em F122M} filter was used to image
\lya\ and the {\em F140LP} for nearby continuum observations, 
redwards of the line. 
 The flux due to continuum processes at the wavelength of the line was then
estimated using the the offline filter combined with various other data. 
 Initially \cite{kun2} we used the archival {\em V}-band images and IUE 
spectra to map the continuum slope (\bet) in the UV to estimate the 
continuum flux by assuming an unbroken power-law continuum between the 
offline and online filters. 
 This enabled us to discuss the morphology of \lya\  in two of our
targets: ESO\,350-IG38 and SBS\,0335-052.

Unfortunately, isolating \lya\ emission/absorption, present difficulties
not normally encountered in narrow band imaging of other spectral lines.   
 Examination of data from another of our targets, \eso338, revealed
some of the flaws in the method previously assumed \cite{hayes}. 
 For this galaxy there is much broadband Wide-Field \& Planetary Camera 2 
(WFPC2) data available, a STIS spectrum of the central regions taken with the 
G140L grating that covers \lya\ as well as the UV spectrum taken with the 
International Ultraviolet Explorer (IUE). 
 We found that when \bmap s created using the {\em ACS/F140LP} and 
{\em WFPC2/F218W} filters were used for the continuum subtraction, the 
galaxy exhibits \lya\ in absorption across the entire starburst region.
 This is in direct contradiction to the STIS spectrum of the same regions; 
the \bmap\ technique caused the subtraction of too much continuum. 
 There are a number of reasons why this simple assumption about \bet\ may 
not hold in this case. 
 Firstly, the IUE spectrum of \eso338\ this target shows that \bet\ 
deviates from a single power-law, flattening out at $\sim1400$\AA. 
 More importantly in this case, the online filter also contains not only 
redshifted \lya\ from the target but a damped Galactic absorption feature
where the high H{\sc i} column density 
($\log(N($H{\sc i}$[$cm$^{-2}]))=20.8$) has removed a large 
fraction of the continuum flux in this filter. 
 Additionally, the presence of discrete stellar \lya\ absorption or emission 
features could lead to erroneous conclusions if not accounted for. 
 
 Clearly some modelling of the continuum slope was necessary in order to 
correctly allow for these features.
 \eso338\ was the ideal target for this experiment due to the large 
amount of existing data against which our results can be tested. 
 We define the Continuum Throughput Normalisation (CTN) factor as the factor
needed to normalise the flux in offline filter to the wavelength of the 
emission line. 
 We used the {\em Starburst99} \cite{lei} synthetic spectra models to 
represent the spectra of the target. 
 These models were chosen due to their inclusion of the latest and most 
appropriate UV stellar atmospheres and cover a large parameter space 
enabling us to test how our selection of parameters may impact upon our 
results. 
 We adopted the full age range (1 to 900Myr) of spectra created with Salpeter
IMF with $M_{\rm up}=100M_{\odot}$, $Z=0.001$, comprising stellar and 
nebular components, and created for an instantaneous burst.
 We obtained the instrument throughput for all the configurations for 
which data were available from the {\sc synphot} package of {\sc iraf/stsdas}.
 By convolving these curves with various synthetic spectra, we could compute
the optical/UV colours and CTN. 
 We adopted the SMC law \cite{prev} to simulate internal extinction in 
\eso338, using the range of $E(B-V)$ from 0.0 to 0.4 magnitudes, 
twice the range previously determined \cite{ost03}.
 For various ages and internal reddenings we looked for any relationships 
between the computed CTN factors and optical/UV colours. 
 The degenerate effects of age and reddening quickly became apparent and no 
single colour (not even \bet) could provide a non-degenerate CTN for a given age and reddening.
 However, for certain {\em pairs} of optical/UV colours, the effects of the 
degeneracy could be disentangled. 
 We determined that to obtain non-degenerate CTNs it is necessary to use one 
colour to trace the UV continuum, which is very sensitive to reddening, and 
one colour that traces the 4000\AA\ discontinuity, which is sensitive to age.  
 We used the {\em F140LP--F336W} and {\em F336W--F439W} colours, 
respectively and thus created a CTN map which we used to subtract the 
continuum from the \lya\ line. 

\begin{figure}
\begin{centering}
\psfig{figure=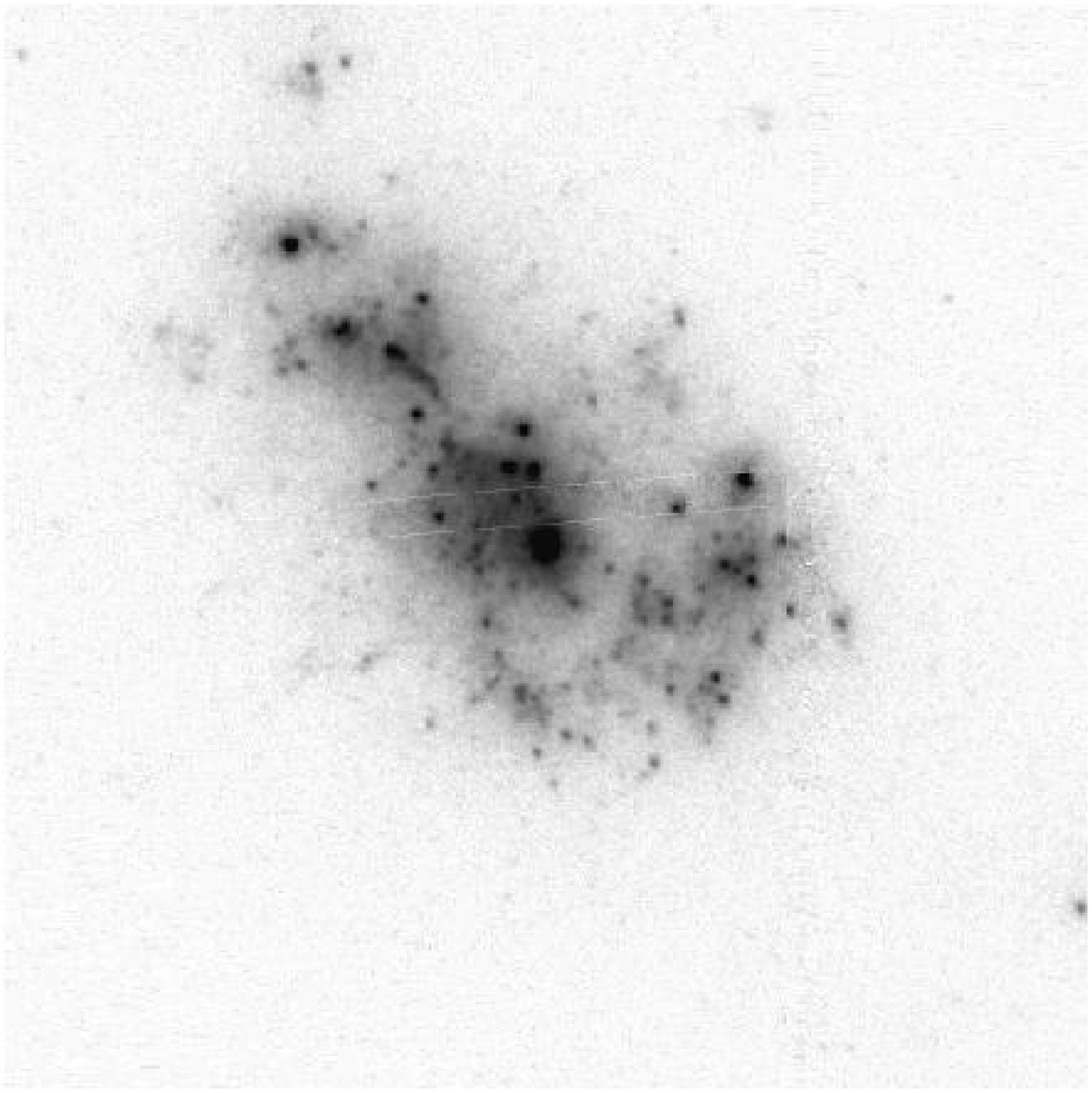,height=2.8in}\hspace{0.5cm}
\psfig{figure=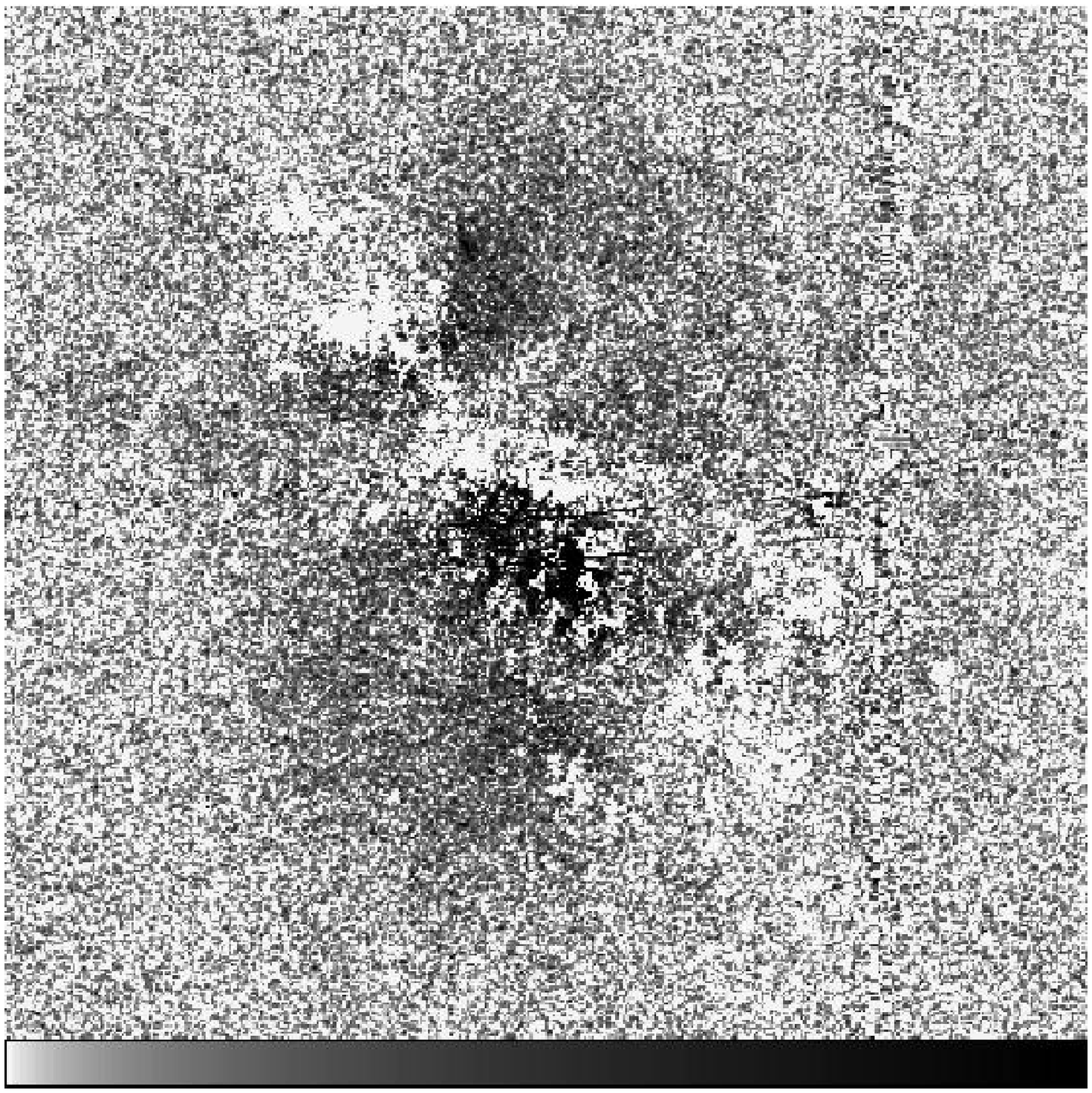,height=2.8in}
\caption{HST/ACS images of \eso338. $15\times15$'' with negative log 
scaling. 
{\em Left}: {\em F140LP} UV continuum. 
{\em Right}: Continuum subtracted \lya\ image. White shows absorption, black
emission.\label{fig:eso338lya}}
\end{centering}
\end{figure}

 Figure \ref{fig:eso338lya} shows the \lya\ line-only image of \eso338.
 Spatially, the emission and absorption features are now fully in agreement 
with those in the two-dimensional STIS spectrum. 
 Emission from the bright central star-clusters is the result of 
outflows in the ISM allowing emission through P-Cygni profiles. 
 In addition, areas of diffuse emission are seen surrounding the central 
starburst, covering a much larger area than the UV continuum emission. 
 
 The \lya\ flux from \eso338\ has been determined from IUE observations to be 
\fluxiuee338. 
 We masked our \lya\ line-only image in a $10\times20$'' ellipse centred on
the brightest central star cluster which matches the size and shape of the 
IUE aperture. 
 Summing the flux in this aperture we measured \fluxacse338; fully consistent 
with the IUE flux within the 10\% photometric accuracy of the instrument. 
 This validates the photometric accuracy of our technique. 

 In the process of selecting the spectra and implementing this method 
we had made assumptions which may have affected our 
results. 
 In order to test this, we ran the software with different sets of spectra,
comparing pixels in the CTN maps against the original assumed model set, and 
computing total fluxes in the continuum subtracted images. 
 By plugging in different sets of model spectra, computing additional spectra,
and modifying the code, tested the effects of: 
metallicities; 
IMFs; 
reddening laws; 
varying stellar and nebular contributions; 
continuous star-formation scenarios; 
nebular emission lines  ({\em Mappings III} \cite{kewl});
and contributions from additional old stellar populations.  
 However when we modified our assumed setup, we obtainted line-only images that
were morphologically indistinguishable by eye.
 Moreover, when reasonable parameter modifications were made, the \lya\ 
fluxes were consistent with the assumed setup to within around 25\%, 
demonstrating that our technique is not highly sensitive to these
assumptions. 

 We compared the total observed \lya\ flux with the total \ha\ 
flux \cite{ost01} of $290\times10^{-14}$\fluxunit. 
 Of this flux, 87\% comes from the regions covered by our ACS observations. 
 Correcting this for Galactic extinction, internal extinction
using the SMC law with $E(B-V)=0.05$ and assuming case A recombination we 
arrive at an implicit \lya\ flux of $3860\times10^{-14}$\fluxunit. 
 Comparing this to our total observed \lya\ flux (corrected for Galactic
extinction), we determine a global escape fraction of 5\%.
 This escape fraction is an order of magnitude lower than that which would 
result if internal extinction were the only destruction mechanism of \lya\ 
photons (44\% for the SMC law with $E(B-V)=0.05$). 

 Another important result of our studies is the discovery of the large areas
of diffuse emission that surround the starburst region. 
 The net \lya\ flux from the central regions is in fact negative and, were it 
not for the diffuse emission regions, \eso338\ would be a net \lya\ absorber. 

\begin{figure}
\psfig{figure=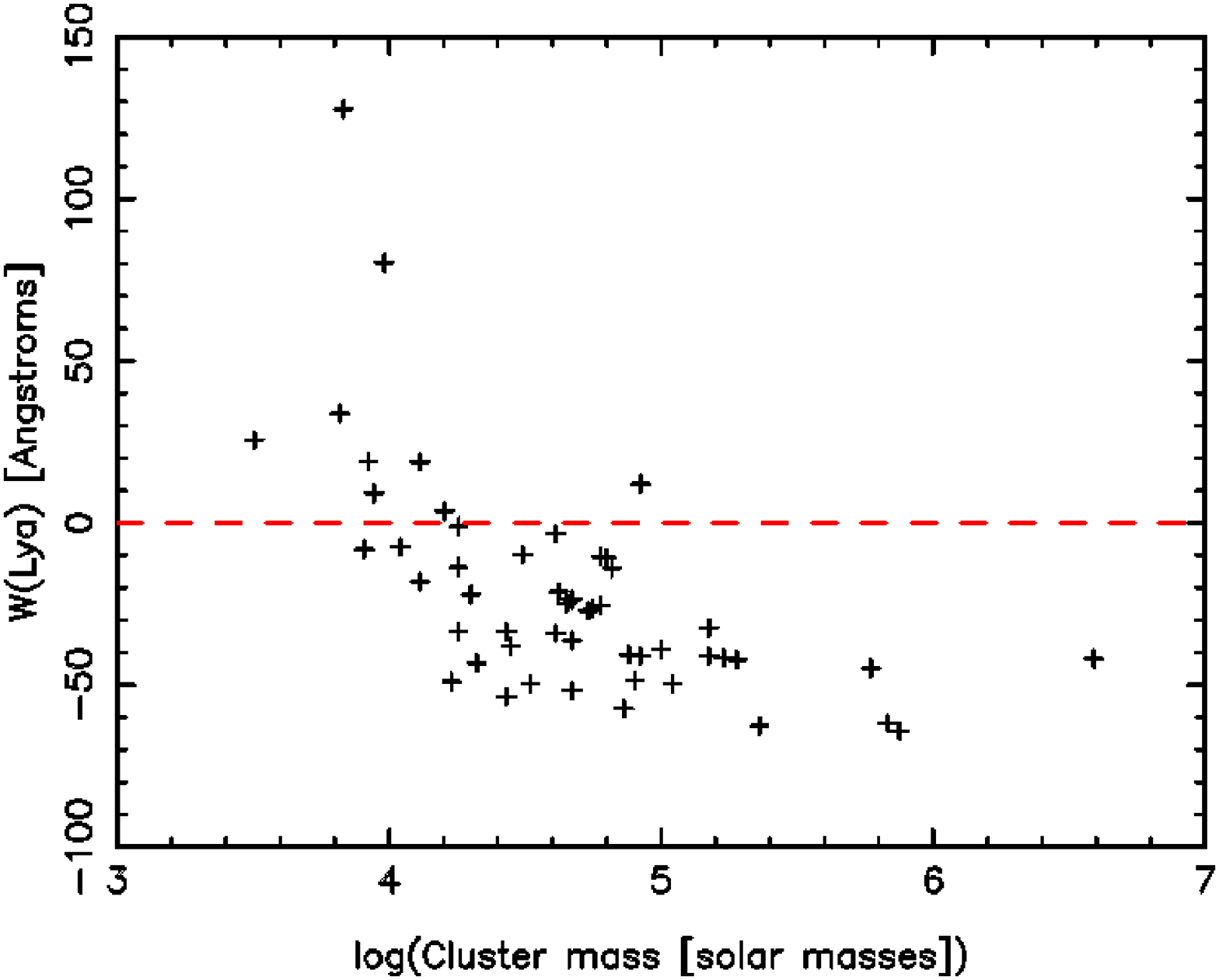,width=3.0in}
\psfig{figure=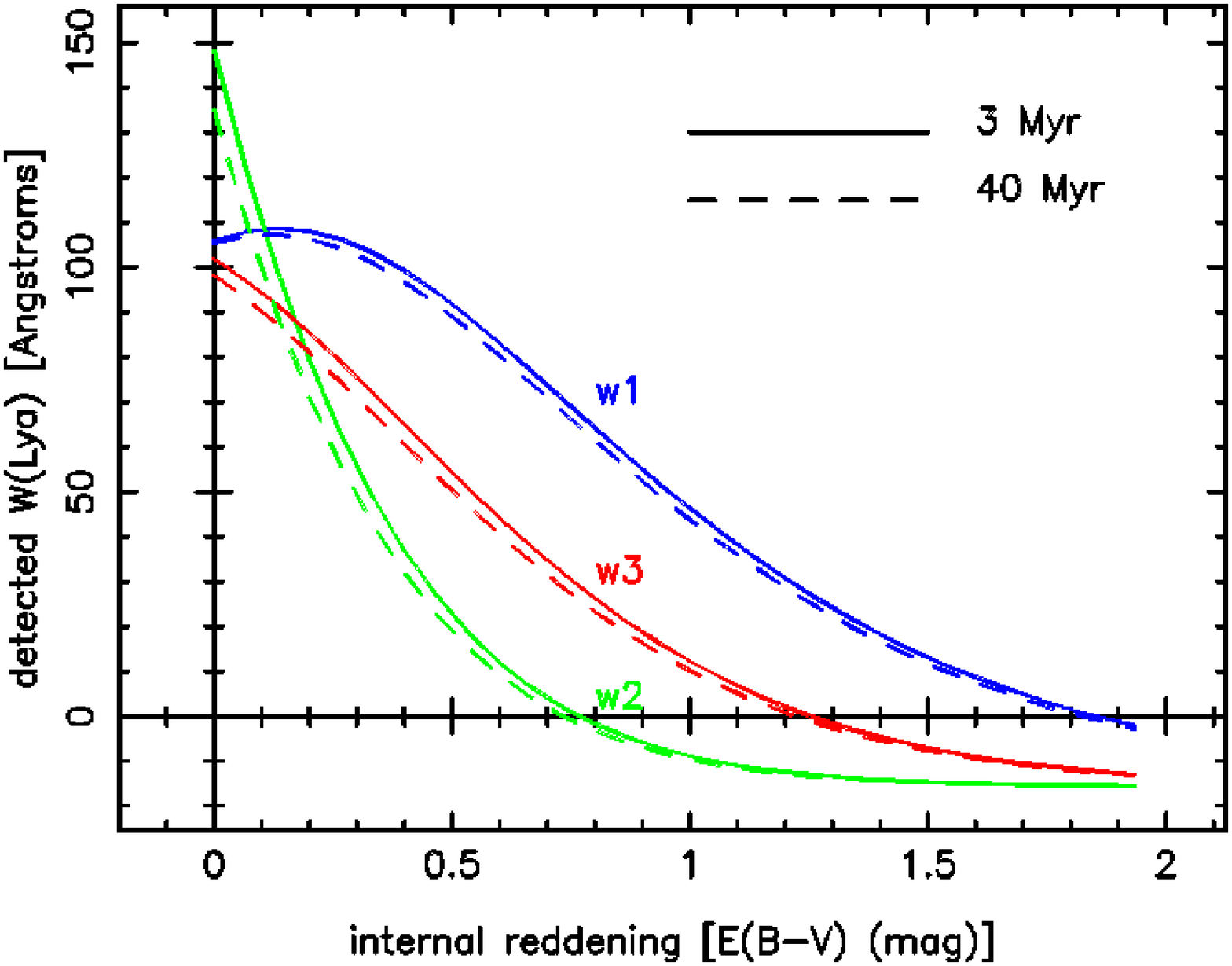,width=3.0in}
\caption{{\em Left}: Plot of \lya\ equivalent width {\em vs.} mass of 
compact star clusters.
{\em Right}: Effects of internal reddening on detected \wlya for a model 
galaxy with \wlya = 100 \AA . 
 The 3 observational techniques described in Sect. \ref{sect:highz} are 
labelled {\em w1}, {\em w2}, and {\em w3} respectively. \label{fig:plot}}
\end{figure}

 In the analysis of our continuum subtracted image, we compared our study
with several other dynamical and kinematic studies. 
 The age, radial distance, mass, colour and luminostiy of compact 
star-clusters \cite{ost03} in \eso338\ were compared
with \lya\ equivalent widths in regions spatially coincident with
the clusters in 2D. 
 The only trend found was between \wlya\ and cluster mass or 
luminosity as can be seen in the left panel of Fig. \ref{fig:plot}. 
 We have not (yet?) found any evidence supporting the prediction that 
\lya\ emission or absorption is function of age. 
 H{\sc i} mapping using the VLA \cite{can} reveals column densities sufficient
to produce damped absorption in all regions of \eso338.
 Fabry-Perot interferrometry \cite{ost99} \cite{ost01} and UVES spectroscopy, 
however, reveal the perturbed and complex nature of the velocity field. 
 Data are suggestive of either a bipolar outflow or a second kinematical
component with centre very close to the central star cluster. 
 The overall outflow velocity is shown to be rather small, indicating that 
the diffuse \lya\ emission is probably the result of small-scale (smaller 
than those resolvable using the VLA) velocity shifts and a clumpy and 
perturbed medium. 

 When experimenting with continuum subtraction techniques, we tested the 
simple assmption of assumed power-law continuum slopes. 
 It is interesting to note that these continuum subtractions produced 
line-only images that {\em were} distinguishable by eye from those produced
by our modelling techniques and in conflict with the STIS spectrum.
 When we assumed a flat continuum we obtained a \lya\ flux three
times greater than that obtained by our modelling and by IUE observations. 
 We would therefore suggest that SED modelling could be a vital for \lya\
observations at low and high redshifts. We are currently in the process
of collecting UV/Optical HST/ACS images for the rest of our sample of local 
galaxies with available HST \lya\ images, which will allow a quantitative
analysis of the kind that we performed for ESO\,338-04.

\section{Detection of \lya\ from high-$z$ objects\label{sect:highz}}

 In an effort to understand the implications and aid the interpretation of 
\lya\ observations, we are currently running simulations of how high- 
(or low-) $z$ \lya\ galaxies appear in narrowband surveys.
 The aim is to estimate uncertainties on derived \lya\ fluxes and equivalent
widths and any biases that may be introduced by astrophysical phenomena or 
observational methods. 
 The prescription is simple: build a model restframe SED with a 
\lya\ emission or absorption line; redshift the SED; compute deteced 
fluxes in a variety of filters and subtract the continuum using various 
commonly employed techniques. 
  
 The {\em Starburst99} spectra are used to which \lya\ features
of arbitrary equivalent width can be added. 
 Internal extinction can then be applied with the various extinction laws 
listed in Sect. \ref{sect:imaging}.
 The SED is then redshifted, placing the galaxy in the distant universe. 
 The effect of intervening H{\sc i} clouds is applied by using random
numbers to generate a distribution of clouds in 
redshift and column density.
 These column densities are converted into equivalent widths at $z=0$ and 
applied to the spectra. 
 Perfect rectangular filters are implemented  with full widths of 50\AA\ 
for the narrowband online filter, and 800\AA\ for the broadband continuum
filters. 
 The online filter is then placed at the observed wavelength of the \lya\ in 
the observed SED under consideration. 
 Three broadband filters are also positioned to measure the continuum: one 
centred at the observed \lya\ wavelength; one 500\AA\ redward of \lya; 
and one 1500\AA\ redward of \lya.
 'Observed' fluxes in these filters are then computed and continuum 
subtractions are perfomed using three methods: 
The first is the method used by most \lya\ searches in which the broadband 
contiuum filter is centred close to \lya. 
 The second method involves using the nearer offline broadband filter to 
measure the continuum 500\AA\ redward of \lya\ and assuming this to represent
the continuum flux at the line. Both these methods assume the continuum level
to be flat over the used wavelength range. 
 In the third  method, we use the flux in the two offline broadband filters
and, assuming a continuous power law continuum, extrapolate to obtain the 
contiuum level at \lya.

 The right panel of Fig. \ref{fig:plot} shows effect of internal 
reddening on the 
observed \wlya\ of a galaxy at $z=2.2$. The restframe SED was built from the 
'standard' setup assumed in our initial studies of \eso338\  (Sect. 
\ref{sect:imaging}) with a restframe \wlya\ of 100\AA.
 Burst ages of 4 and 30 Myr were used with the SMC law used to describe
internal reddening.
 It shows how \lya\ is removed by relatively modest extinction although 
interestingly it demonstrates how, at very low extinction, \wlya\ is 
{\em overestimated} by commonly used observational techniques. 
 Moreover, we also find some evidence that in these surveys, the mean
detected \wlya\ may be artificially increased by up to 20\% of its 
restframe value by the 
presence of intervening intergalactic clouds. 
 The interpretation for this being that there is a high probability for clouds
to remove flux from the blue end of the broadband filter, leaving the line 
itself unaffected. These results for the high-$z$ simulations will be presented 
in a forthcoming paper
(Hayes et al. 2005, in preparation).


\end{document}